# Fiber modal noise mitigation by a rotating double scrambler

G. Raskin*[a], J. Pember[b], D. Rogozin[ab], C. Schwab[b], D. Coutts[bc]
[a] Institute of Astronomy, KU Leuven, 3001 Leuven, Belgium
[b] Department of Physics and Astronomy, Macquarie University, NSW 2109, Australia
[c] MQ Photonics Research Centre, Macquarie University, NSW 2109, Australia

## ABSTRACT

Fiber modal noise is a performance limiting factor in high-resolution spectroscopy, both with respect to achieving high signal-to-noise ratios or when targeting high-precision radial velocity measurements, with multi-mode fiber-fed high-resolution spectrographs. Traditionally, modal noise is reduced by agitating or "shaking" the fiber. This way, the light propagating in the fiber is redistributed over many different modes. However, in case of fibers with only a limited number of modes, e.g. at near-infrared wavelengths or in adaptive-optics assisted systems, this method becomes very inefficient. The strong agitation that would be needed stresses the fiber and could lead to focal ratio degradation, or worse, to damaging the fiber. As an alternative approach, we propose to make use of a classic optical double scrambler, a device that is already implemented in many high-precision radial-velocity spectrographs, to mitigate the effect of modal noise by rotating the scrambler's first fiber end during each exposure. Because of the rotating illumination pattern of the scrambler's second fiber, the modes that are excited vary continuously. This leads to very efficient averaging of the modal pattern at the fiber exit and to a strong reduction of modal noise. In this contribution, we present a prototype design and preliminary laboratory results of the rotating double scrambler.

## 1. INTRODUCTION

In high-precision radial-velocity (RV) spectroscopy, multi-mode optical fibers are the work-horse technology to relay the light from the telescope focal plane to the entrance of the spectrograph. They offer some clear advantages, mainly with respect to instrument stability:

- Optical fibers provide mechanical isolation between telescope and spectrograph, avoiding gravitational flexure and vibration;
- Mounting the instrument away from the telescope makes it possible to house it in an isolated and controlled environment, improving thermal and pressure stability;
- Transmission efficiency of modern optical fibers is very high over the entire optical wavelength range;
- Finally, optical fibers *scramble* the image at the fiber entrance, this way strongly reducing the effect of seeing and guiding variations on the illumination of the spectrograph.

However, optical fibers also come with two important drawbacks. Firstly, they produce *focal ratio degradation* (FRD), the tendency to broaden the cone angle of the transmitted beam. Micro-bending, imperfections and stress in the fiber scatter the propagation angles of the rays of light that are guided in the fiber, resulting in a larger aperture of the output beam or in other words, a degraded focal ratio. This will result in loss of throughput or reduced spectral resolution, unless the spectrograph beam diameter and the size of the spectrograph optics are increased.

A second drawback of multi-mode optical fibers is that they suffer from modal noise [1]. Modal noise refers to the noise caused by the variable spatial distribution of monochromatic light at the output face of a multi-mode optical fiber. This irregular speckle pattern (Figure 1) is very sensitive to any disturbance of the fiber (e.g. fiber position, stress, temperature, injection geometry, wavelength), making it highly unstable. Astronomical spectra cover a wide wavelength range but in case of a high-resolution spectroscopy, each narrow spectral line can be considered as a monochromatic source, prone to modal noise. The effect of modal noise increases when the number of modes guided in the fiber is reduced. The number of modes $N$ in a multi-mode fiber is proportional to $N \sim (d/\lambda)^2$, with $d$ the fiber diameter and $\lambda$ the wavelength of the guided light. Consequently, the impact of modal noise on near-infrared observations or on adaptive-optics assisted spectroscopy with narrow multi-mode fibers is much more pronounced [2][3] compared to visible light observations through large fibers with a typical diameter of ~ 100 μm.

---

* gert.raskin@kuleuven.be

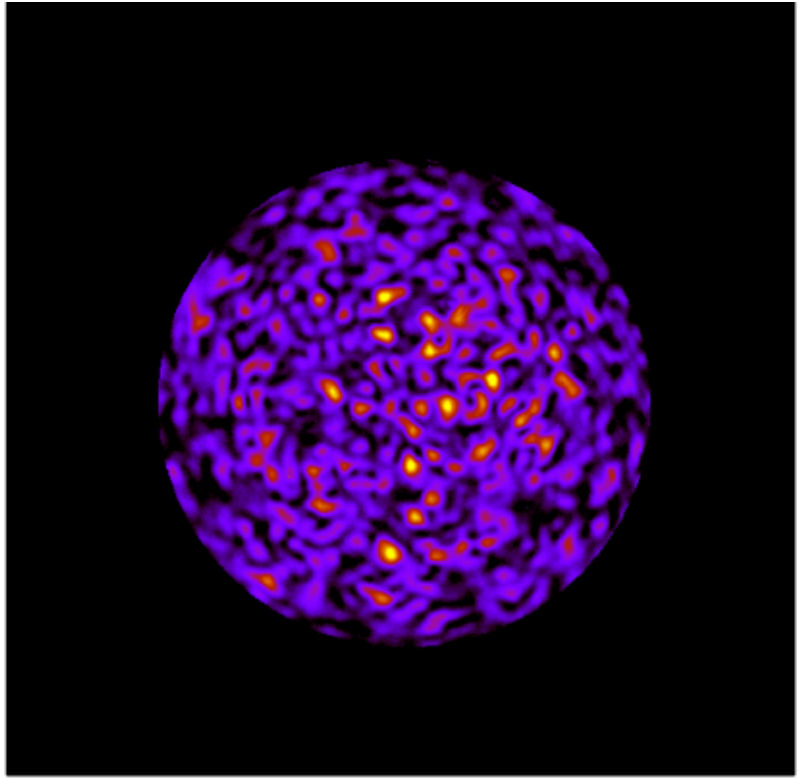

Figure 1. False-color image of the spatial distribution of the light at the exit face of a static fiber (60 μm core diameter) illuminated with monochromatic (λ = 633 nm) light. The speckle pattern is caused by the modal nature of guiding the light through the fiber.

We distinguish two aspects of modal noise that are relevant for astronomical spectroscopy. Firstly, modal noise affects the photometric accuracy of the observation [4]. This is especially noticeable in the case of beam truncation, e.g. by a slit or because of undersized spectrograph optics, leading to mode filtering. This will limit the maximum achievable signal-to-noise ratio that can be obtained by the fiber-fed instrument [5]. Secondly, modal noise distorts the spectral line profiles, resulting in less accurate determination of the central wavelengths of these spectral lines, and consequently, in an error on the radial velocity measurement.

Various techniques have been described to mitigate modal noise in astronomical spectroscopy [6]. By far the most popular approach is agitating or *shaking* the fiber at a frequency that is substantially higher than the integration time of the exposure. Many mechanical fiber agitators are in use, operating at various agitation amplitudes, some of them using combined systems that provide non-harmonic actuation e.g. [5][7][8][9][10]. Another way of mechanical agitation consists of stretching the fiber [11]. Dynamical optical diffuser are very efficient in mitigating modal noise too, especially in combination with an integrating sphere, but their low throughput only allows them to be used for calibration light sources like a laser frequency comb [12]. In this paper, we present a mitigation solution for the presence of modal noise that is based on the variation of the illumination of the fiber input, that way continuously varying the modes that are excited in the fiber and averaging out the effects of modal noise.

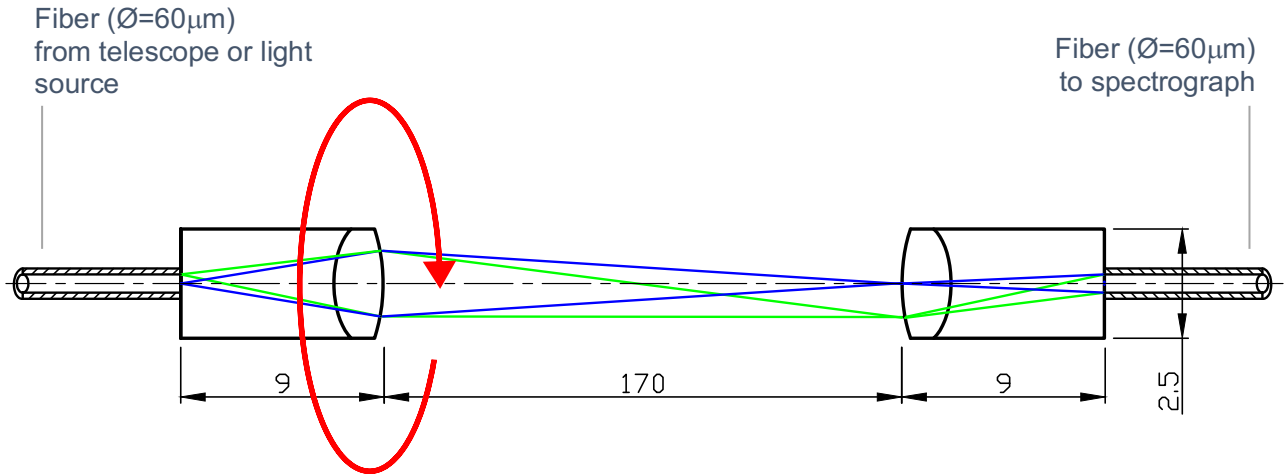

Figure 2. Layout, dimensions (vertical scale is doubled for clarity) and ray-trace of an optical double scrambler. The left part of the optical system can continuously rotate in order to mitigate the effect of modal noise.

## 2. A ROTATING DOUBLE SCRAMBLER

Optical fibers are extremely good at scrambling the spatial information at the entrance of the fiber. However, scrambling of angular information is much less efficient. As a consequence of incomplete scrambling, seeing and telescope guiding fluctuations will affect the stability of the illumination of the spectrograph, resulting in a degraded precision of the radial-velocity measurements. Double optical scramblers are routinely used to increase the scrambling performance of the fiber link and to improve the illumination stability of high-precision radial-velocity spectrographs [13][14][15]. These scramblers consist of two identical lenses, connected to the fiber faces at a junction in the fiber train (Figure 2). The first lens images the first fiber face on the pupil of the second part and vice-versa. This way, the near and far field images of both fiber parts are swapped, and scrambling of spatial as well as angular information is greatly increased. Modal noise, unfortunately, is not reduced by this type of static optical double scrambling [5].

A more efficient alternative to mechanic fiber agitation is varying the input illumination of the fiber. To achieve this, we equipped the first part of the optical double scrambler with a rotation mechanism. To avoid fiber wind-up, we continuously rotate between -180°±15° and +180°±15°. The nominal rotation speed is 30°/s, leading to 12 s for one full rotation. The random offset of ± 15° was added to avoid any systematic effects of the short moment of slower rotation due to acceleration and deceleration always taking place at the same angular position. The second part of the scrambler, connected to the fiber leading to the spectrograph is fixed. For stability reasons, the rotating scrambler is mounted vertically. Both fiber ends and lenses are mounted on 5-axis adjustable mounts (XYZ/Tip/Tilt) for aligning of the optical and rotation axes. The rotating fiber is 3 m long and left free hanging to avoid mechanical stress from the rotation. The complete setup is shown in Figure 3.

Correct alignment of the rotating scrambler is a precarious aspect of our method and requires an iterative approach because of the effect of continuous rotation. Firstly, the optical axis of the rotating lens has to aligned with the rotation axis of the mechanism, requiring 4 degrees of adjustment. Secondly, the optical axis of the fixed lens is aligned with the rotating lens and the focus is adjusted (5 degrees of freedom). Especially in the case of narrow fibers and small lenses, precise and stable alignment, and a wobble-free rotation mechanism are required to avoid throughput losses.

For our preliminary laboratory experiments, we recovered the decommissioned double scrambler of the HERMES spectrograph, described in [16]. This scrambler is equipped with two achromatic lenses ($F = 6$ mm), cemented on two 60-μm circular fibers (length > 6 m). We illuminate the telescope fiber with an $F$/4 beam from a 633-nm monochromatic light source. A 40X microscope objective images the speckle pattern at the exit of the spectrograph fiber on a CMOS camera. We use an integration time of one minute in our experiments, to average out the effects of modal noise over five revolutions of the rotation stage.

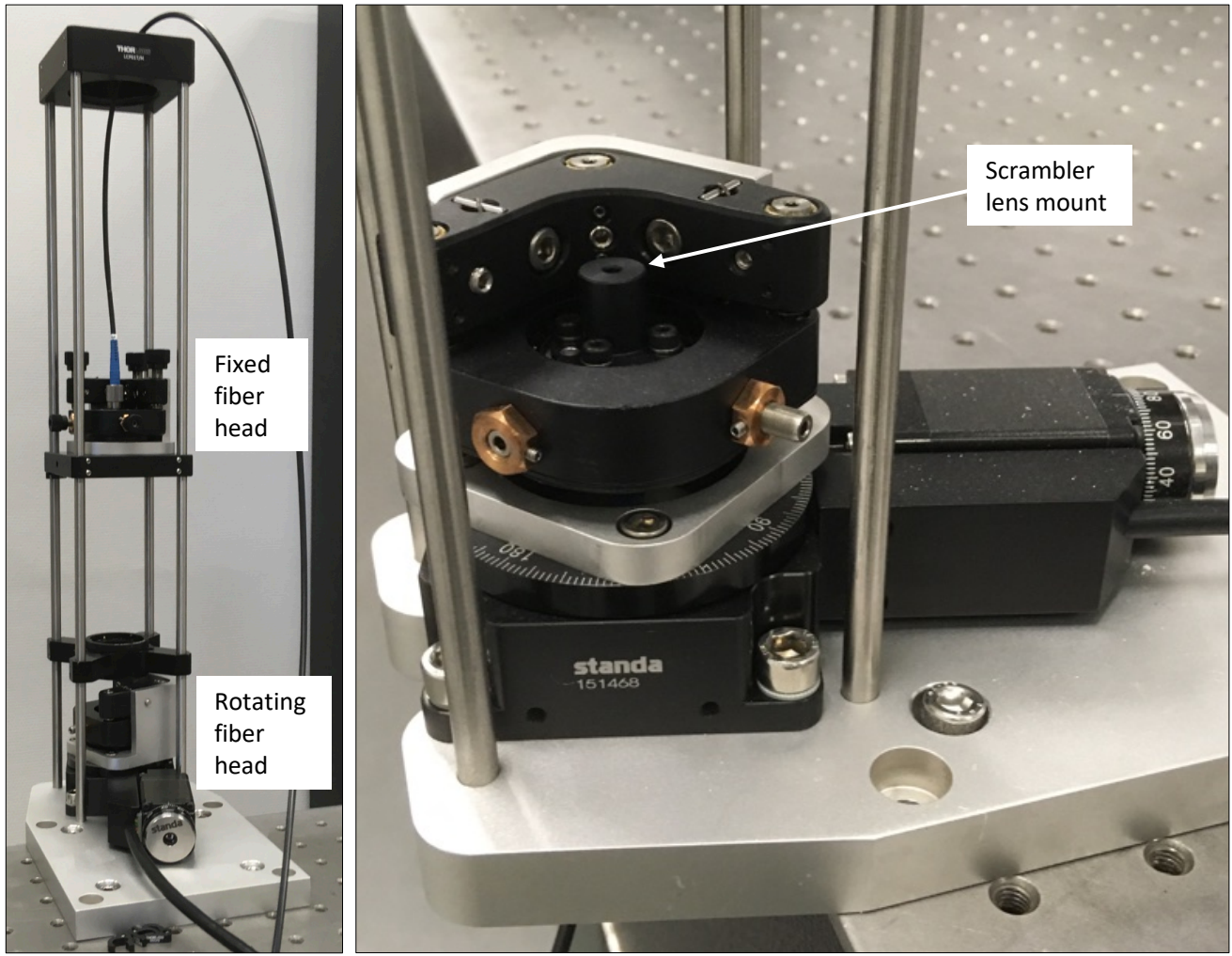

Figure 3. Left: picture of the complete rotating scrambler setup, the illuminated fiber (coming from the telescope) is connected to the rotating fiber head, the fiber at the fixed fiber head feed the light to the spectrograph. Right: Close-up view of the rotation stage and the XYZ/Tip/Tilt alignment mount.

## 3. RESULTS

The impact of the rotating scrambler on modal noise can be immediately appreciated in the microscope images of the fiber exit face without (Figure 1) or with (Figure 4) rotation. Without scrambler rotation, the modal speckle pattern is very pronounced while it is almost completely averaged out during a 1-minute integration over five 360° rotations of the scrambler. In Figure 5, we plot the azimuthally averaged power spectral density of these fiber face images. At spatial frequencies corresponding with typical modal speckle size (∅ 2 μm speckle ≈ 15 camera pixels), the rotating scrambler reduces the structure in the image by almost two orders of magnitude.

Changes in the speckle pattern also affect the centroid position of the fiber image, and hence, the position of the spectral line. This results in an error on the radial velocity measurement. To assess the effect of modal noise and of the rotating scrambler on the RV accuracy, we took a series of measurements, for each data point we modified the position and bend radius of the input fiber in order to create a different input illumination pattern. Subsequently, we measured the centroid position of the image at the exit fiber and scaled the scatter on the output image centroids to a RV error obtained with a typical RV spectrograph. The results are plotted in Figure 6. Illuminated with white light, hence in the absence of modal noise, this results in a centroid scatter of 1.4 m/s, roughly corresponding with the measurement accuracy of our setup. Illuminated with monochromatic light ($\lambda = 633$ nm), this increases to 18 m/s without rotating scrambler. Rotating the scrambler results in a reduction to 1.6 m/s, hardly above the measurement noise floor. These scatter values correspond with the centroid error in spectral direction for a single spectral line at a resolution of $R = 100\,000$. The global radial velocity error due to modal noise for a complete spectrum, containing thousands of absorption lines, would be reduced to the 10 cm/s level.

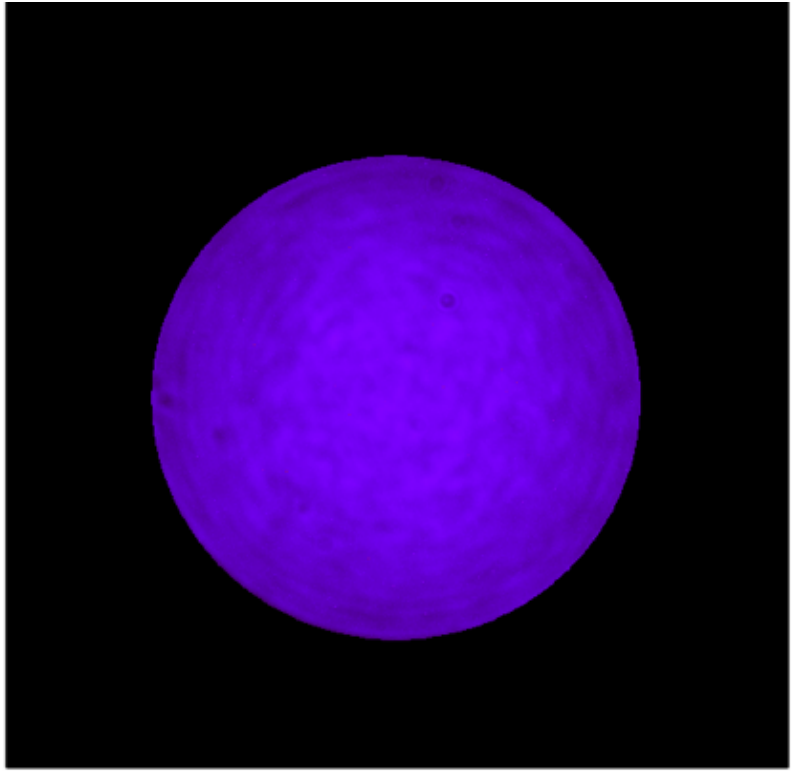

Figure 4. False-color image of the spatial distribution of the light at the exit face of the same fiber as shown in Figure 1 (with exactly the same color scale), averaged over five scrambler revolutions (integration time of one minute).

Alignment of the rotating optical double scrambler is a tricky aspect of this setup. Both the images and the pupils of the two fiber faces need to stay precisely aligned during the rotation to avoid light loss. We measured a peak-to-peak flux variation of 5% over a complete revolution, indicating there is still room for a small alignment improvement. We estimate the average light loss due to the remaining alignment errors at 3% approximately. Clearly, the mechanical design of the rotation mechanism is critical as the correct alignment needs to be maintained over extended periods and over many rotation cycles.

## 4. CONCLUSION

The solution that we presented for mitigating the effects of modal noise in multi-mode optical fibers can be generally applied to any high-resolution spectrograph that is equipped with an optical double scrambler. It is extremely efficient for both stellar spectra as well as for highly coherent wavelength calibration sources like the laser frequency comb. It is especially attractive obtaining extreme-precision radial velocity measurements in the near-infrared where the number of fiber modes is limited because of the longer wavelengths and mechanical fiber agitation becomes insufficient of stressful for the fiber. Furthermore, adaptive-optics assisted systems could benefit from this method too as they typically use narrow optical fibers that, just like the near-infrared fibers, also guide only a very limited number of modes. In that case however, the exigencies on the optical alignment and the rotation mechanism become even tighter because of the smaller fiber dimensions.

Finally, as a more elegant alternative to rotating one fiber end back and forth continuously, we envisage to install a dove prism on a rotation stage centered between the lenses on both fiber ends that can now remain completely fixed. This would allow smoother operation as well as a constant rotation speed at the cost of more delicate alignment and of a small throughput loss because of the Fresnel reflections at the prism surfaces.

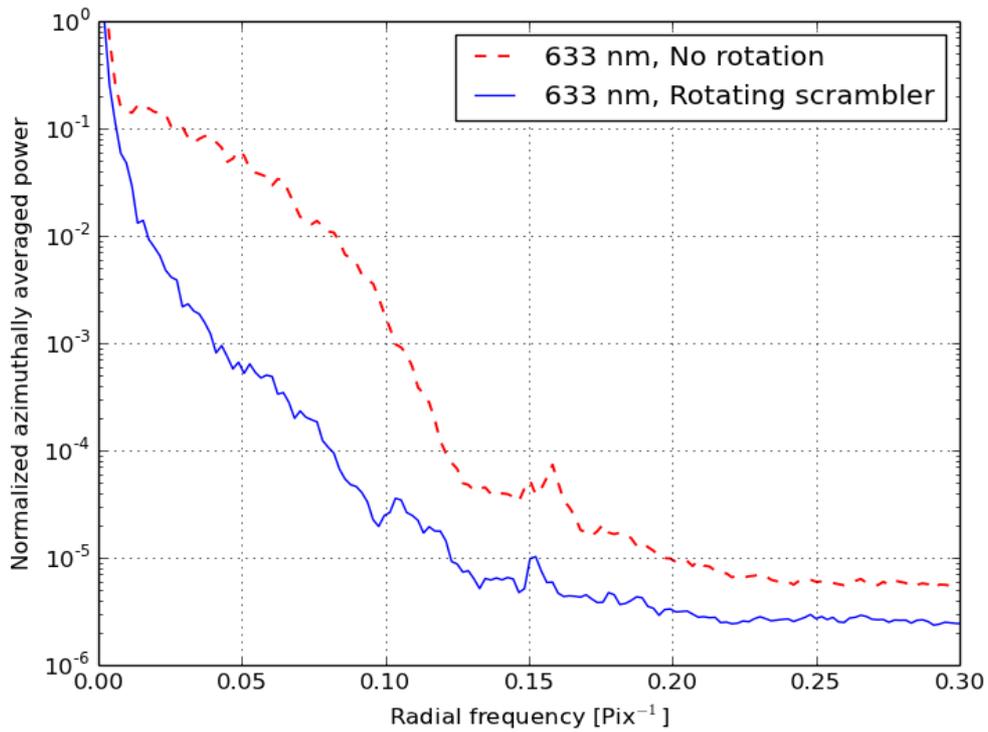

Figure 5. Azimuthally averaged power spectrum of the images of the fiber exit face shown in Figure 1 (no scrambler rotation) and Figure 4 (with rotation).

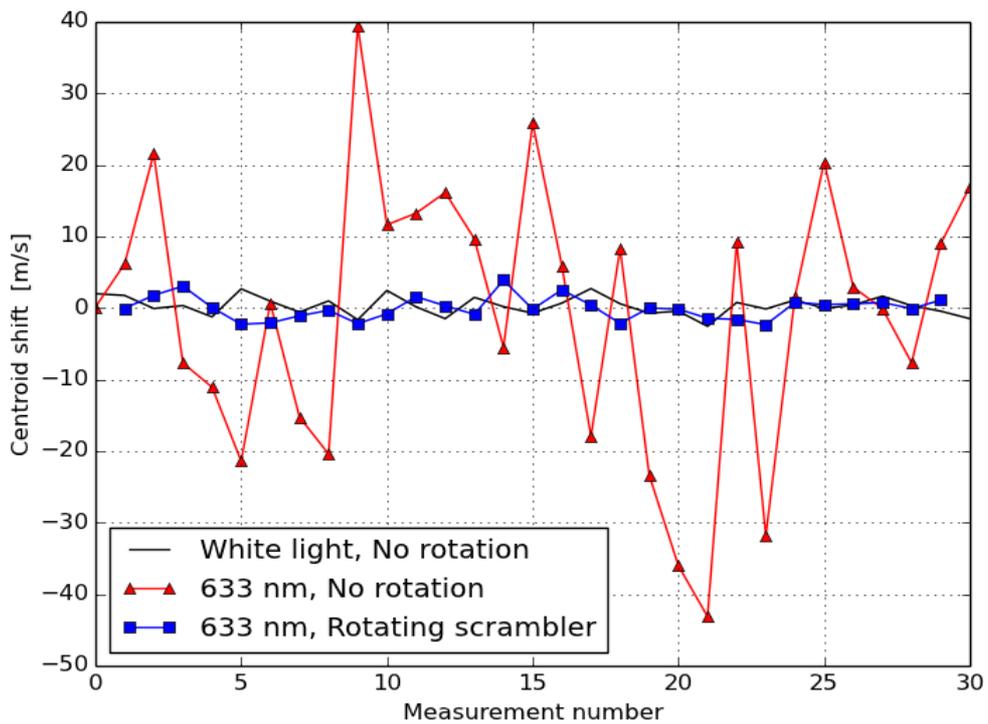

Figure 6. Radial velocity scatter of a single spectral line with and without scrambler rotation (white light illumination is shown as noise floor reference of this measurement).

## ACKNOWLEDGMENTS

GR acknowledges support from the Fund for Scientific Research of Flanders (FWO) under grant I002519N (International Research Infrastructure program).

## REFERENCES

[1] E. G. Rawson, J. W. Goodman, and R. E. Norton, 'Frequency dependence of modal noise in multimode optical fibers', *JOSA*, vol. 70, no. 8, pp. 968–976, Aug. 1980, doi: 10.1364/JOSA.70.000968.

[2] F. Wildi *et al.*, 'NIRPS: an adaptive-optics assisted radial velocity spectrograph to chase exoplanets around M-stars', in *Techniques and Instrumentation for Detection of Exoplanets VIII*, San Diego, United States, Sep. 2017, p. 41, doi: 10.1117/12.2275660.

[3] K. S. McCoy, L. Ramsey, S. Mahadevan, S. Halverson, and S. L. Redman, 'Optical fiber modal noise in the 0.8 to 1.5 micron region and implications for near infrared precision radial velocity measurements', in *Society of Photo-Optical Instrumentation Engineers (SPIE) Conference Series*, Sep. 2012, vol. 8446, pp. 84468J-84468J, doi: 10.1117/12.926287.

[4] U. Lemke, J. Corbett, J. Allington-Smith, and G. Murray, 'Modal noise prediction in fibre spectroscopy - I. Visibility and the coherent model', *Mon. Not. R. Astron. Soc.*, vol. 417, no. 1, pp. 689–697, Oct. 2011, doi: 10.1111/j.1365-2966.2011.19312.x.

[5] J. Baudrand and G. A. H. Walker, 'Modal Noise in High-Resolution, Fiber-fed Spectra: A Study and Simple Cure', *Publ. Astron. Soc. Pac.*, vol. 113, pp. 851–858, Jul. 2001, doi: 10.1086/322143.

[6] R. O. Reynolds and A. Kost, 'A comparison of methods for the reduction of fiber modal noise in high-resolution spectrographs', in *Society of Photo-Optical Instrumentation Engineers (SPIE) Conference Series*, Jul. 2014, vol. 9151, pp. 91514K-91514K, doi: 10.1117/12.2057024.

[7] A. Roy, S. Halverson, S. Mahadevan, and L. W. Ramsey, 'Scrambling and modal noise mitigation in the Habitable Zone Planet Finder fiber feed', in *Society of Photo-Optical Instrumentation Engineers (SPIE) Conference Series*, Jul. 2014, vol. 9147, pp. 91476B-91476B, doi: 10.1117/12.2055342.

[8] D. P. Sablowski, D. Plüschke, M. Weber, K. G. Strassmeier, and A. Järvinen, 'Comparing modal noise and FRD of circular and non-circular cross- section fibres', vol. 1, no. 1, pp. 0–1, 2015.

[9] M. M. Sirk *et al.*, 'A optical fiber double scrambler and mechanical agitator system for the Keck planet finder spectrograph', in *Ground-based and Airborne Instrumentation for Astronomy VII*, Austin, United States, Jul. 2018, p. 234, doi: 10.1117/12.2312945.

[10] C. Frank, F. Kerber, G. Ávila, N. Di Lieto, G. Lo Curto, and A. R. Manescau Hernandez, 'A fiber scrambling unit for the laser frequency comb of ESPRESSO', in *Ground-based and Airborne Instrumentation for Astronomy VII*, Austin, United States, Jul. 2018, p. 244, doi: 10.1117/12.2313484.

[11] N. Blind, U. Conod, and F. Wildi, 'Few-mode fibers and AO-assisted high resolution spectroscopy: coupling efficiency and modal noise mitigation', 2017, [Online]. Available: http://arxiv.org/abs/1711.00835.

[12] S. Mahadevan, S. Halverson, L. Ramsey, and N. Venditti, 'Suppression of Fiber Modal Noise Induced Radial Velocity Errors for Bright Emission-Line Calibration Sources', *Astrophys. J.*, vol. 786, no. 1, pp. 18–18, 2014, doi: 10.1088/0004-637X/786/1/18.

[13] T. R. Hunter and L. W. Ramsey, 'Scrambling properties of optical fibers and the performance of a double scrambler', *Publ. Astron. Soc. Pac.*, vol. 104, pp. 1244–1251, Dec. 1992, doi: 10.1086/133115.

[14] T. M. Brown, 'High precision Doppler measurements via echelle spectroscopy', vol. 8, pp. 335–344, 1990.

[15] M. Casse and F. Vieira, 'Comparison of the scrambling properties of bare optical fibers with microlens coupled fibers', in *Optical Telescopes of Today and Tomorrow*, Mar. 1997, vol. 2871, pp. 1187–1196, [Online]. Available: http://esoads.eso.org/abs/1997SPIE.2871.1187C.

[16] G. Raskin *et al.*, 'HERMES: a high-resolution fibre-fed spectrograph for the Mercator telescope', *Astron. Astrophys.*, vol. 526, pp. A69–A69, Dec. 2010, doi: 10.1051/0004-6361/201015435.